\documentclass[twocolumn,pra,aps,showpacs,flushbottom]{revtex4-1}
\usepackage{xspace,amsmath,amsfonts,amsthm,amssymb,amsbsy,graphicx,color}

\begin{document}

\newtheorem{theo}{Theorem} \newtheorem{lemma}{Lemma}

\title{The approach to typicality in many-body quantum systems}

\author{Shawn Dubey, Luciano Silvestri, Justin Finn, Sai Vinjanampathy, and Kurt Jacobs}
\affiliation{Department of Physics, University of Massachusetts at Boston, 100 Morrissey
Blvd, Boston, MA 02125, USA 
}

\begin{abstract}
The recent discovery that for large Hilbert spaces, almost all (that is, \textit{typical}) Hamiltonians have eigenstates that place small subsystems in thermal equilibrium, has shed much light on the origins of irreversibility and thermalization. Here we give numerical evidence that many-body lattice systems generically approach typicality as the number of subsystems is increased, and thus provide further support for the eigenstate thermalization hypothesis. Our results indicate that the deviation of many-body systems from typicality decreases exponentially with the number of systems. Further, by averaging over a number of randomly-selected nearest-neighbor interactions, we obtain a power-law for the atypicality as a function of the Hilbert space dimension, distinct from the power-law possessed by random Hamiltonians. 
\end{abstract}

\maketitle

\section{Introduction}

In the last few years tremendous progress has been made in understanding how irreversibility and thermalization emerge from the reversible evolution of quantum mechanics~\cite{Gemmer03, Popescu06, Goldstein06, Gemmer06, Reimann08, Rigol08, Gemmer09, Linden09, Reimann10, Goldstein10a, Goldstein10b, Riera11, Banuls11}. Recall that traditional statistical mechanics requires three basic assumptions, that together imply that the equilibrium state of a small subsystem is the Boltzmann state~\cite{Riera11, Landau}. Let us call the small subsystem the \textit{system}, and the large system the \textit{universe}. The three assumptions are: i) The state of the universe is a uniform mixture over all its accessible states, and thus all states within some small energy band; ii) Over the energy range of the system, the density of energy levels of the rest of the universe (often called the \textit{bath}) increases exponentially with energy, and iii) that the coupling of the system to the rest of the universe is small on the scale of the energy levels of the system (weak coupling). 

The modern understanding removes the need for the first, unjustified and rather odd assumption, that the state of the universe is mixed. It turns out that \textit{almost every} pure state of the universe, within a given energy band, gives the \textit{same} state for the system (when the rest of the universe is traced out) as the uniformly mixed state on the same energy band (note that this uniformly mixed state is the  \textit{micro-canonical} state for that energy band). When almost all members of a large set have some property in common, these members are referred to as \textit{typical}~\footnote{To be precise, an element $\omega$ of a set $\Omega$ is referred to as being typical when the following are true: i) The number of elements in $\Omega$ is $N$; ii) the element $e$ shares properties with a set of states of size $N'$; iii) the ratio $N'/N$ tends to unity as $N\rightarrow\infty$. Thus the statement ``almost all states are typical states'', is true when $N$ is sufficiently large.} This ``canonical typicality''~\cite{Popescu06, Goldstein06} property of pure states provides a compelling reason as to why the equilibrium state of a large isolated system is obtained correctly by assuming it is mixed (namely that it is the micro-canonical ensemble). Since almost all pure states are typical, the universe can be expected to spend almost all its time in these states, and the system will therefore remain close to the micro-canonical state as far as every small subsystem is concerned. 

The above result is not the end of the story. The typicality of pure states implies that for large (high-dimensional) vector spaces, almost all Hermitian operators (and thus almost all potential Hamiltonians) also have a typicality property~\cite{Goldstein10a, Goldstein10b}. Specifically, if we select a Hamiltonian ``at random'' by choosing its eigenstates randomly and independently from the Haar measure, then almost all Hamiltonians will have typical eigenstates. This is the typicality property of Hamiltonians. It is also worthwhile to note that if almost all Hamiltonians are typical under one ``reasonable'' measure, then we can expect them to be typical under other ``reasonable'' measures. While this conclusion is far from rigorous --- we have not defined the word ``reasonable'' --- we can nevertheless expect that in general ``random'' Hamiltonians will have typical eigenstates, even if their eigenstates are not picked precisely from the Haar measure. 

The concept of Hamiltonian typicality is important for the following reason. If the Hamiltonian of a many-body system is typical, and its eigenvalues are sufficiently non-degenerate, then along with a few additional, easily justifiable conditions (e.g. the individual subsystems are weakly interacting)~\cite{Goldstein10a, Riera11}, the many-body system will thermalize. By ``thermalize" we mean that the density matrix of every small subsystem relaxes to a steady-state, and this steady-state is the Boltzmann state. This result was proved only recently by Riera, Gogolin, and Eisert \textit{et al.}~\cite{Riera11}), while prior to this, strong arguments along these lines were given by Golstein \textit{et al.}~\cite{Goldstein10a, Goldstein10b} (see also the closely-related papers by Riemann~\cite{Reimann08, Reimann10}). Note that we have not provided any evidence yet that many-body Hamiltonians \textit{are} typical, but we will turn to this question shortly. 

The reason that typical Hamiltonians lead to thermalization can be outlined fairly simply. Firstly, the eigenstates of a typical Hamiltonian are typical, meaning that if the universe is in any one of them, every small subsystem will be in the state predicted by the micro-canonical ensemble for the universe. In fact, it had previously been conjectured independently by Deutsch~\cite{Deutsch91} and by Srednicki~\cite{Srednicki94} that the Hamiltonians of large systems have this property, and this conjecture is now termed the \textit{eigenstate thermalization hypothesis}, or ETH~\cite{Rigol08, Rigol12}. The next step is to examine what happens when we write the initial state of the universe in the basis of the universe eigenstates. If the initial state of the universe is $|\psi(0)\rangle = \sum_n c_n |E_n\rangle$, where the states $|E_n\rangle$ are the energy eigenstates, then the density matrix of the universe at time $t$ is 
\begin{eqnarray}
   \rho_{\mbox{\scriptsize U}}(t)  & = &   \sum_n |c_n|^2 |E_n\rangle \langle E_n| \nonumber  \\  & + & \sum_{n\not=m} c_n c_m^* |E_n\rangle \langle E_m| \exp[-i(E_n - E_m)t/\hbar].   \nonumber 
   \label{eq:unev}
\end{eqnarray}
We now note that because the energy gaps, $\Delta_n = E_n-E_{n-1}$ are non-degenerate, the off-diagonal terms --- those with $n\not= m$, will mutually dephase, and since each term is sinusoidal, and has zero mean, these terms can potentially cancel each-other out. Specifically, if we take the trace over all but one subsystem, so as to obtain the density matrix of that subsystem, then we have 
\begin{eqnarray}
   \rho(t)  & = &   \sum_n |c_n|^2  \mbox{Tr}_{\mbox{\scriptsize b}} \left[ |E_n\rangle \langle E_n| \right]  \nonumber  \\  & + & \sum_{n\not=m} c_n c_m^*  \mbox{Tr}_{\mbox{\scriptsize b}} \left[ |E_n\rangle \langle E_m| \right] \exp[-i(E_n - E_m)t/\hbar] ,   \nonumber 
   \label{eq:unev}
\end{eqnarray}
where $\mbox{Tr}_{\mbox{\scriptsize b}} [\cdots]$ denotes the trace over the ``bath'' part of the universe. Since each off-diagonal term is time-dependent, for the subsystem to reach a steady-state, the partial traces of the off-diagonal elements \textit{must} cancel themselves out. Showing that they do requires considerable technical detail~\footnote{That small subsystems of a many-body system with non-degenerate energy gaps do reach a thermal equilibrium has been proven by Linden \textit{et a.}~\cite{Linden09} and Riera \textit{et a.}~\cite{Riera11}. But determining the \textit{timescale} of thermalization is still a largely open problem~\cite{Linden09, Hutter11}.}, but if we assume that it \textit{does} reach a steady-state, then this steady-state \textit{must} be given by the diagonal terms alone. Now, every eigenvector $|E_n\rangle$ gives the same density matrix for the system, and this is $\rho_{\mbox{\scriptsize micro-can}}$, being the state that results from starting with the micro-canonical state, and tracing out the rest of the universe. So we have  
\begin{eqnarray}
   \rho_{\mbox{\scriptsize ss}} = \lim_{t\rightarrow\infty}\rho(t)  & = &   \sum_n |c_n|^2  \mbox{Tr}_{\mbox{\scriptsize b}} \left[ |E_n\rangle \langle E_n| \right] \nonumber \\ 
    & = &  \mbox{Tr}_{\mbox{\scriptsize b}} \left[ \rho^{\mbox{\scriptsize universe}}_{\mbox{\scriptsize micro-can}}  \right] . 
   \label{eq:unev}
\end{eqnarray}
It now follows from the usual arguments of statistical mechanics that $\rho_{\mbox{\scriptsize ss}}$ will be equal to the Boltzmann state so long as 1) each subsystem interacts sufficiently weakly with the rest of the subsystems (the bath), and 2) the density of the energy levels of the bath increases exponentially with energy. For many-body systems this second assumption follows from the combinatorics of combining many identical systems together, so long as the coupling between the systems is small (perturbative) compared to the gaps between their individual energy levels~\cite{Riera11}. 


To summarize, if the Hamiltonian of a many-body system is typical, and its energy spectrum has sufficiently non-degenerate energy gaps, then it will thermalize. This thermalization will be micro-canonical, and will also be canonical if the couplings between the subsystems are sufficiently weak.  This suggests the following explanation for the ubiquity of thermalization in nature: since almost all Hermitian operators for large systems are typical, it is very likely that the Hamiltonian of any specific many-body system will be typical, and thus will thermalize. We suggest, however, that there may be a piece missing from this picture. As we discussed above, if one picks a Hamiltonian using some randomization procedure, then indeed it can be expected to be typical. But the Hamiltonians of many-body systems are not obtained in this way. They are built in a simple and ordered manner, usually by short range, few-body interactions between simple systems, where the interactions are dictated by an often highly symmetric grid. To obtain a full understanding of the origin of thermal behavior, ideally one would show that the procedure that generates many-body Hamiltonians typically results in Hamiltonians that are both typical and have highly non-degenerate energy gaps. 

We do not attempt to prove here that many-body Hamiltonians are generically typical. But motivated by this question, we obtain numerical results showing how a chain of spin systems approaches typicality as its length is increased. To do this we define a measure of the typicality of a Hamiltonian, and examine how this scales with the dimension of the relevant Hilbert space.   We compare this with the scaling of the typicality exhibited by random Hamiltonians as their size increases. We further provide numerical evidence, using a chain of spin-1 systems, that this approach to typicality is generic --- that is, that it occurs for almost all interaction Hamiltonians that one can choose for the nearest-neighbor spin-spin coupling. We note that previous numerical evidence for the ETH, and thus the typicality of many-body Hamiltonians, has been presented in~\cite{Rigol08, Rigol10}. Prior to this numerical analysis had shown that non-integrable spin systems have a typical eigenvalue distribution (being the \textit{Wigner-Dyson} distribution, see below)~\cite{Poilbla93, Hsu93}. Earlier pioneering work on the thermodynamic behavior of spin chains is given in~\cite{Jensen85}. 

Before we launch into our numerical analysis, we feel it is worth noting a few more things regarding typicality and many-body systems. Firstly, it is telling that random Hamiltonians have been shown to correctly reproduce global aspects of the eigenvalue distributions of many-body systems (see, e.g.~\cite{Mehta04,Guhr98}). If indeed the Hamiltonians of many-body systems are typical in various respects, then this would account for the above fact. The procedure used to obtain the random Hamiltonians in question is to select each of their independent matrix elements independently from a zero-mean Gaussian probability density. If the matrix elements are purely real, then the resulting ``ensemble'' of Hamiltonians  is called the Gaussian orthogonal ensemble, or GOE.  (There is a similar procedure for selecting Hamiltonians with complex elements, but the GOE is sufficient for our purposes --- it gives all Hamiltonians that are time-reversal invariant). Not only are the eigenvectors of  GOE Hamiltonians typical, but their eigenvalues are highly non-degenerate. This is the famous ``energy-level repulsion'', in which the distribution of the gaps between adjacent energy levels is small for small gap-size. The distribution of the adjacent gaps is called the Wigner-Dyson distribution. The fact that the eigenvalues of almost all GOE Hamiltonian's do follow this single distribution implies that this is a ``typical'' property of GOE Hamiltonians. Thus GOE Hamiltonians have stronger typicality that the Hamiltonians we have been discussing above, for which it is only the eigenvectors that are chosen at random. Experimental investigations of nuclear (many-body) systems have confirmed that they also follow the Wigner-Dyson distribution~\cite{Gurevichand57}, and thus share this typicality property. Finally we note that the GOE ensemble also has the important property that it is invariant under all orthogonal transformations~\cite{Mehta04}. In the following, when we use the term ``random Hamiltonian'', we will always mean a Hamiltonian selected from the GOE ensemble. 


The question of the typicality of many-body systems is made more curious by the fact that some non-integrable many body systems do not have typical Hamiltonians. This occurs when the Hamiltonians of each of the subsystems varies substantially from system to system (an example of this is a spin-1/2 lattice immersed in a randomly varying magnetic field)~\cite{Gornyi05, Vadim07, Znidaric08, Ioffe10, Berkelbach10, Aleiner10, Pal10, Gogolin11}. In this case the eigenstates of a lattice system are localized on the lattice, a phenomena first suggested by Anderson in his seminal paper on localization~\cite{Anderson58}. Though less relevant in this present context, there is another, more special class of systems that do not thermalize: systems that contain conserved quantities in additional to the energy. In this case the eigenstates still tend to be typical, but now they are typical states within subspaces whose states are degenerate with respect to all the conserved quantities. Such systems therefore still relax to a steady-state, but it is not the Boltzmann state, but a ``generalized'' Boltzmann state that takes into account the additional conserved quantities~\cite{Rigol07}. Integrable systems are an extreme case of this, in that they have the same number of conserved quantities as the number of subsystems~\cite{Chowdhury04, Girardeau60, Lieb63, Andrei80}. Because they have so many conserved quantities, the conserved subspaces are too small for typicality to play any role, and as a result they do not usually settle down to a steady-state at all (although they sometimes do thermalize for certain initial conditions~\cite{Grisins11} and exhibit diffusive behavior~\cite{Langer09, Steinigeweg10}).  

In the following section we define a measure of the ``atypicality'' of a Hamiltonian, along with a quantity that indicates the extent to which a Hamiltonian is degenerate. In Section~\ref{numres} we describe our example systems and present the numerical results. Section~\ref{conc} concludes.  

\section{A measure of atypicality} 

Consider a many-body system consisting of $N$ qubits, in which the energy gap between the excited and ground states of each qubit is $\hbar \Delta$. Because we are interested in thermalization, we are free to make the coupling strength between each pair of qubits, $\hbar g$, as small as we wish. In the limit of small $g$ ($g \ll \Delta$), the spin-spin interaction couples only states that are degenerate with respect to the free Hamiltonian, and thus states in which the same number of spins are pointing up. (This degeneracy of the free Hamiltonian should not be confused with the degeneracy of the full Hamiltonian --- as explained above, the latter must not have a significant fraction of degeneracies if the system is to thermalize.) As is usual, we will refer to these degenerate subspaces as sectors. It follows that thermalization happens \textit{separately} on each sector. The Hamiltonians that generate the thermalization are therefore the sub-blocks of the interaction Hamiltonian corresponding to each sector. If the spin chain has $N$ spins, then we will denote the sector in which $M$ spins are up as $N_M$, and the interaction Hamiltonian on this space as $H_{N,M}$. We will further write the eigenstates of $H_{N,M}$ as $|n\rangle_{N,M}$. 

Let us denote the state of the universe that is maximally mixed on the sector $N_M$ by $\Xi_{N,M}$. Define the state of a single spin, given by tracing out all other spins from the state $\Xi_{N,M}$ as $\rho_{N,M}$. We will denote the elements of $\rho_{N,M}$ by $\rho_{N,M}^{(i,j)}$, where $i$ and $j$ are ``up" or ``down". The population of the spin-up state is therefore $\rho_{N,M}^{(\mbox{\scriptsize up},\mbox{\scriptsize up})}$. One can easily determine $\rho_{N,M}$ merely by counting the number of basis states of $N_M$ for which a single chosen qubit is in the up-state, and dividing this by the total dimension of the space. The dimension of $\Xi_{N,M}$ is ``$N$ choose $M$'', given by $N!/(M!(N-M)!)$, and the number of states in which a chosen qubit is in the up-state is ``$N-1$ choose $M-1$''. The micro-canonical ensemble therefore implies that the population of the up-state of every qubit is  
\begin{equation}
      \rho_{N,M}^{(\mbox{\scriptsize up},\mbox{\scriptsize up})} =  \frac{\mbox{$N$ choose $M$}}{\mbox{$(N-1)$ choose $(M-1)$}} = \frac{M}{N} . 
\end{equation}
The off-diagonal elements of $\rho_{N,M}$ are zero, and of course $\rho_{N,M}^{(\mbox{\scriptsize down},\mbox{\scriptsize down})} = 1 - M/N$. 
 
The property of $H_{M,N}$ in which we are interested, is the extent to which each of its eigenstates places every spin in the state $\rho_{N,M}$. Let us denote the $n^{\mbox{\scriptsize\textit{th}}}$ eigenstate of the universe, in the sector $N_M$, by $|E_n\rangle_{N,M}$. Note that there are ``$N$ choose $M$" eigenstates in this sector. The state of the $j^{\mbox{\scriptsize\textit{th}}}$ qubit, when the universe is in this eigenstate, is 
\begin{equation}
   \rho_{N,M}(n,j)  =  \mbox{Tr}_j \left[ |E_n\rangle_{N,M} \langle E_n |_{N,M} \right] , 
\end{equation} 
where $\mbox{Tr}_j [\cdots]$ denotes the trace over all but the $j^{\mbox{\scriptsize th}}$ qubit. A measure of the deviation of the density matrix $\rho_{N,M}(n,j)$ from that predicted by the micro-canonical ensemble is given by the difference between the populations they predict for the spin-up state. If we denote the spin-up state by $| \mbox{up}\rangle$, then this difference is 
\begin{eqnarray}
   \delta_{j,n} & = &  \langle \mbox{up} | \rho_{N,M}(n,j)  | \mbox{up} \rangle -  \langle \mbox{up} |  \rho_{N,M} | \mbox{up} \rangle \nonumber \\ 
    & = & \langle \mbox{up} | \rho_{N,M}(n,j)  | \mbox{up} \rangle - \frac{M}{N}  . 
\end{eqnarray} 
A root-mean-square measure of the typicality of $H_{M,N}$ is obtained by averaging the square of this  difference over all the spins and all the eigenstates in the sector $N_M$. Since there are ``$N$ choose $M$" eigenstates in this sector, this rms measure is 
\begin{equation}
     \delta_{\mbox{\scriptsize rms}} =  \sqrt{ \frac{M!(N-M)!}{N! N} \sum_{n,j} \delta_{j,n}^2 }  . 
\end{equation} 
We will refer to $\delta_{\mbox{\scriptsize rms}}$ as the \textit{atypicality} of the Hamiltonian in the given sector. 

We must also verify that $H_{M,N}$ does not contain a significant fraction of degenerate eigenvalues, and that this remains true as $N$ increases. This is because the eigenvectors are only well-defined by the Hamiltonian if their corresponding eigenvalues are non-degenerate, and thus the question of typicality is meaningless if there is significant degeneracy. This is paralleled by the fact that degeneracy of the eigenvalues implies immediately degeneracy of the energy gaps, and thus destroys the connection between typicality and thermalization. To quantify the fraction of degenerate eigenvalues, we introduce the following rather crude measure, which is sufficient for our purposes. We calculate all the energy gaps between adjacent energy levels. We then calculate the average value of the energy gap, which we denote by $\langle \varepsilon \rangle$, and define a threshold $\varepsilon_{\mbox{\scriptsize thresh}} = \langle \varepsilon \rangle/10$. Finally, we determine the maximum number of energy gaps whose combined sum is less than $\varepsilon_{\mbox{\scriptsize thresh}}$. This maximum number, divided by the total dimension of the space, we will call the ``degeneracy fraction'', and denote it by $f_{\mbox{\scriptsize deg}}$. 

\section{Numerical results}
\label{numres}
We use as our first numerical example a one-dimensional lattice of spin-1/2 systems (a chain of qubits), with nearest-neighbor interactions.  The Hamiltonian for this many-body system (which we will also refer to as ``the universe'') is  
\begin{equation}
     H = \hbar \Delta \sum_{n=1}^N  \sigma_{z}^{(n)} + \hbar g \sum_{j=1}^{N-1} \sigma_j\otimes \sigma_{j+1} , 
     \label{eq::sH}
\end{equation}
where $\sigma_z^{(n)}$ is the Pauli $z$-operator for the $n^{\mbox{\scriptsize\textit{th}}}$ spin, and $\sigma_j$ is the interaction operator for spin $j$. All the interaction operators are identical, and chosen so that the system is non-integrable. We set the interaction operators to be $\sigma_j = \cos\theta \sigma_z + \sin\theta\sigma_x$ with $\theta = 0.375\pi$.    

We wish to examine how the rms deviation from typicality, $\delta$, as well as the degeneracy fraction, $f_{\mbox{\scriptsize deg}}$, changes as the size of the spin chain is increased, and we do this by fully and exactly diagonalizing the Hamiltonian on a single sector for up to 17 spins. It turns out, somewhat remarkably, that for sectors in which exactly half the spins are pointing up (that is, $M = N/2$) all the eigenstates are \textit{perfectly} typical (that is, $\delta = 0$) for any number of spins (at least up to $N = 16$). Presumably this is a result of a symmetry in this case. However, thermalization requires \textit{all} the sectors to have typical Hamiltonians, and so we examine $\delta$ for four other sectors as we increase the length of the chain. The first three sectors are those with $M = \mbox{trunc}(N/2)$ (for $N$ odd), $M = \mbox{trunc}(N/2)-1$, and $M = \mbox{trunc}(N/2)-2$, where the function $\mbox{trunc}(\cdot)$ discards the fractional part of its argument. The last sector we examine is that with $M$ fixed at the value $6$ as $N$ is increased.

\begin{figure}[t]
\leavevmode\includegraphics[width=1.0\hsize]{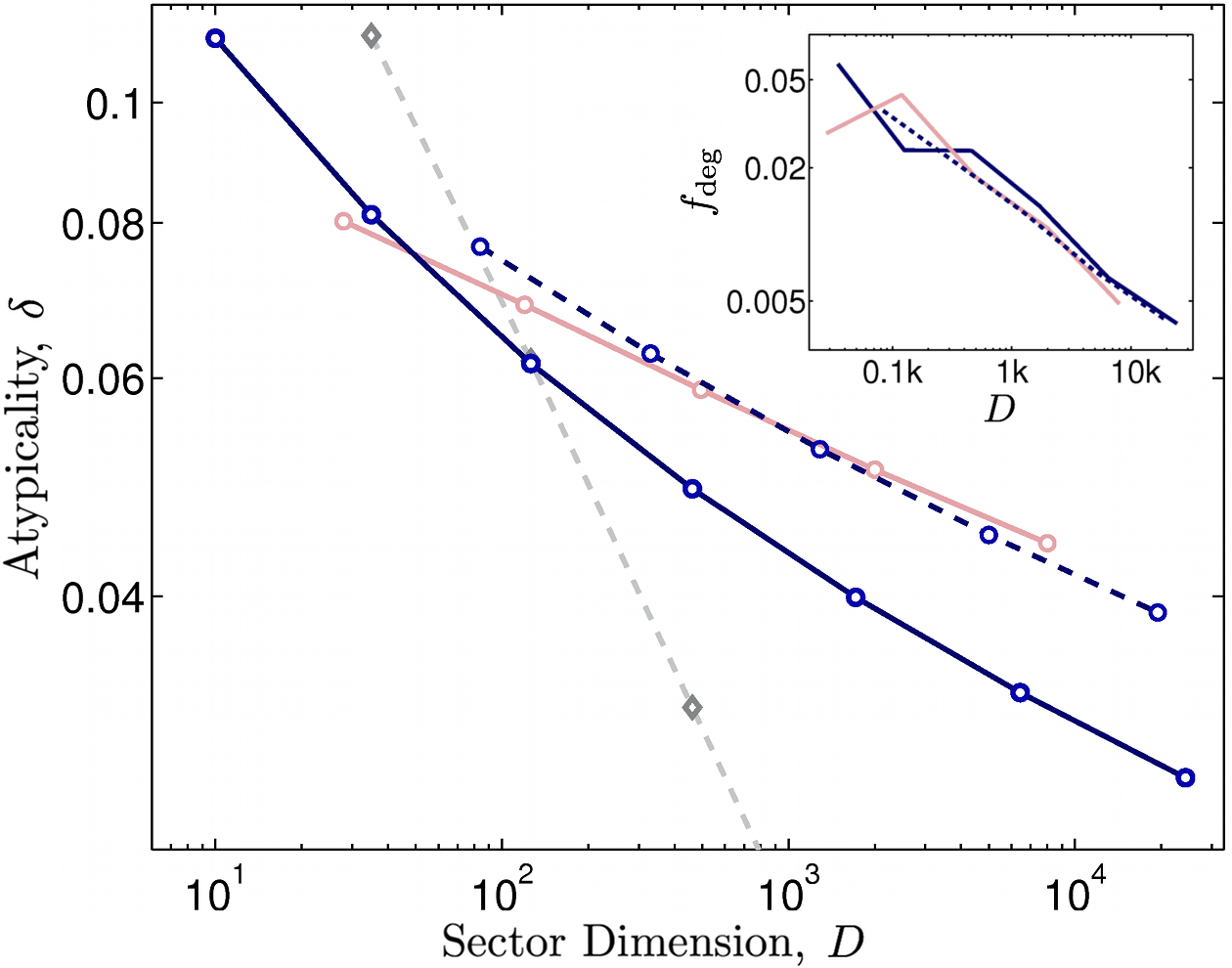}
\caption{(color online) Here we show the deviation from typicality of various sectors of a spin-1/2 chain, as the length of the chain is increased, and compare this to that for random (GOE) Hamiltonians of the same size. On the $x$-axis we plot the dimension of the sector. Dark-solid line:  if the chain has $N$ spins, then the sector has $M = \mbox{trunc}(N/2)$ spins in the up-state, where the function $\mbox{trunc}(\cdot)$ discards the fractional part of its argument (only odd values of $N$ are plotted for these sectors --- see text); dark-dashed line: the sector has $M = \mbox{trunc}(N/2)-1$ spins in the up-state; light-solid line: the sector has $M = \mbox{trunc}(N/2)-2$ spins in the up-state; light-dashed line: the deviation from typicality of random Hamiltonians as a function of the dimension. The inset gives the corresponding ``degeneracy fraction" for the sectors (see text), as a function of their dimension.} 
\label{fig1}
\end{figure}
  
In Fig.~\ref{fig1} we display a log-log plot of $\delta$ for our first three sectors as the length of the chain, $N$, is increased. The value on the $x$-axis is the dimension of the sector, $D$, which increases exponentially with $N$ for large $N$. In the same figure we plot the value of $\delta$ for  random Hamiltonians as a function of their dimension. For random Hamiltonians, the value of $\delta$ fluctuates somewhat from one sample Hamiltonian to another, as one would expect. One also expects these fluctuations to decrease as the dimension of the sector increases. To reduce the fluctuations we average $\delta$ over 50 samples, and it is this average value, $\langle \delta_{\mbox{\scriptsize ran}} \rangle$ that we show in Fig.~\ref{fig1}. The behavior of $\langle \delta_{\mbox{\scriptsize ran}} \rangle$ settles down quickly to a power law as the dimension increases. From the analyses in~\cite{Gemmer06, Gemmer09, Reimann07}, one expects that the deviation for random Hamiltonians will scale asymptotically as $1/sqrt{D}$, and our numerical results confirm this. 

From Fig.~\ref{fig1} we see that the atypicality decreases monotonically for the three sectors, although it has not settled into a power law. The degeneracy fraction also decreases with $D$. Since the curvature of $\log(\delta)$ verses $\log(D)$ is positive, it is not clear whether the atypicality continues to decrease without bound as $D\rightarrow\infty$. What is clear is that $\delta$ does not decrease as fast as it does for random Hamiltonians.  

The behavior of the atypicality for sectors in which $M$ is fixed at a small value, while $N$ increases, is quite different from that for the three sectors we examined above where both $M$ and $N$ increase. We have calculated $\delta$ for sectors with $M=6$, and $N$ running from 13 to 20. For $N=13$ the sector has size $D = 1716$, and for $N = 20$, $D = 38,760$. However in this case $\delta$ does not decrease with increasing size: the values we obtained were, for increasing $N$, $\delta = (399, 415, 448, 441, 441, 434, 434, 418)\times 10^{-2}$. This is certainly interesting, but since both $N$ and $M$ are large for many-body systems at any appreciable temperature, it poses no problem for the ETH.  

\begin{figure*}[t]
\leavevmode\includegraphics[width=1.0\hsize]{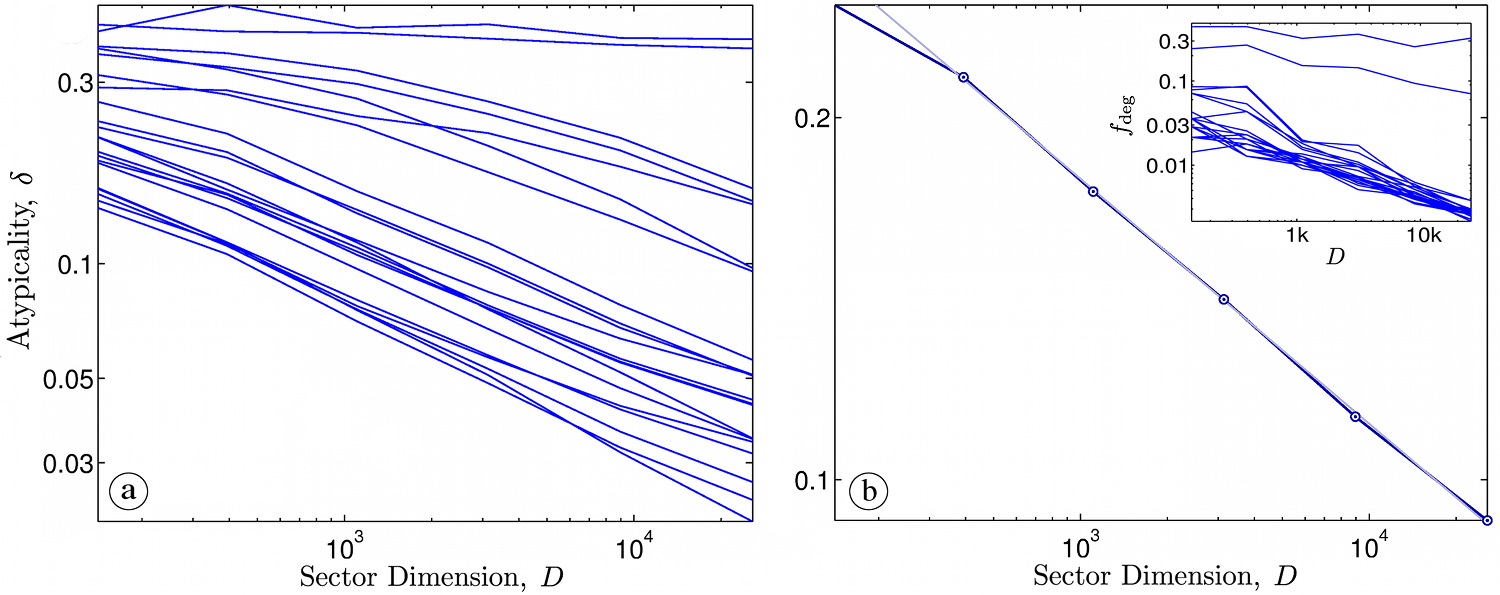}
\caption{(color online) The deviation from typicality of a specified sector (see text) of a nearest-neighbor spin-1 chain, measured by a single real number $\delta$. As the length of the chain is increased, the dimension of the specified sector also increases, and $\delta$ is plotted here as a function of this dimension, $D$. (a) Here we plot $\delta$ for 21 randomly-chosen nearest-neighbor interactions. (b) The average value of $\delta$, $\langle \delta \rangle$, averaged over the 21 different nearest-neighbor interactions. The circles (connected by the dark line) give the data-points for $\langle\delta\rangle$. The light line is the (least squares) best-fit straight line through all but the left-most data-point for $\langle\delta\rangle$. This indicates that $\langle\delta\rangle$ decreases as $D^{-1/5}$. The inset gives the ``degeneracy fraction", also as a function of the sector dimension, for the 21 nearest-neighbor interactions.} 
\label{fig2}
\end{figure*}

We also wish to know whether many-body systems approach typicality for almost all choices of the two-body interaction. To this end we now consider chains of spin-1 systems (qutrits), as this will allow us to explore a more extensive range of interaction Hamiltonians. For a lattice of spin-1/2 systems, there is very little freedom in choosing the interaction, especially if we demand that this interaction operator is a product of identical operators for each system. For a spin-1 chain each system now has three energy levels, which we choose to be equally spaced in energy, and the interaction operator for each system is a 3-by-3 matrix. The Hamiltonian for this chain is 
\begin{equation}
     H = \hbar \Delta \sum_{n=1}^N  J_{z}^{(n)} + \hbar g \sum_{j=1}^{N-1} A_j\otimes A_{j+1} , 
\end{equation}
where $J_z^{(n)}$ is the $z$-component of the spin-1 angular momentum operator for the $n^{\mbox{\scriptsize\textit{th}}}$ spin, and $A_j$ is the interaction operator for the $j^{\mbox{\scriptsize\textit{th}}}$. All the interaction operators $A_j = A$ are identical for every spin. 

The reason we choose the energy levels for every spin-1 system to be equally spaced is to provide degenerate sectors, just as for the spin-1/2 chain. Since the three energy levels are $-\Delta, 0$, and $\Delta$, for a chain of $N$ systems we will examine the degenerate subspace with energy equal to $0$ (as this is the largest degenerate subspace). This time the diagonal elements of the ideal thermal density matrix have two degrees of freedom, so we include both in our measure of typicality. That is, in calculating $\delta$, we sum the squared-differences for two of the diagonal elements. 

We restrict ourselves to interaction operators, $A$, that are zero on the diagonal, and this leaves us with the freedom to chose the three real off-diagonal elements. (In fact since the overall scaling is unimportant, the space of the interaction operators is two-dimensional.) We proceed by picking a value for $A$ from the GOE, and calculate $\delta$ for the sector with energy $0$, for chains containing 6 to 11 spins. In Fig.~\ref{fig2}(a) we show $\delta$ as a function of the sector dimension for 21 randomly sampled interactions $A$. This shows that for the majority of these interactions, the atypicality decreases monotonically with the length of the chain. In the inset in Fig.~\ref{fig2}(b), we show the degeneracy fraction as a function of the sector dimension, and this also decreases as the dimension increases. 

We now average over the atypicality, $\delta$, for all 21 samples shown in Fig~\ref{fig2}(a), and this average is plotted in Fig.~\ref{fig2}(b). After the first data point, the average value of $\delta$ appears to decrease as a power of the sector dimension (a straight line on the log-log plot), and so we determine the least-squares best-fit line though the data-points. This line is also shown in Fig.~\ref{fig2}(b), and gives the power law as $\delta \propto D^{-\alpha}$, where $\alpha = 0.204 \pm 0.003$.

\section{Conclusion}
\label{conc}
We have shown, for nearest-neighbor chains of both qubits and qutrits, that as the length of the chain increases, the Hamiltonians within the degenerate sectors of the non-interacting spins (those responsible for thermalization) become increasingly typical. For a chain of spin-1 systems, we found a clear indication that the atypicality, averaged over a set of choices for the nearest-neighbor interactions, scaled as a power-law in the dimension of the sector. This provides considerable support for the conjecture that many-body systems are generically typical, and thus for the eigenstate thermalization hypothesis. 

The exponent with which the atypicality scales for the nearest-neighbor spin-1 chain is not the same as that for random (GOE) Hamiltonians. Perhaps further investigations of the way in which atypicality scales for many-body systems will help to shed light on the origin of typicality in these systems, and whether this typicality can be understood as resulting from their structure. Finally, we note that since the dimension of the degenerate sectors scales exponentially with the length of the chain, our results indicate that the atypicality also decreases exponentially with the chain length. 

\section*{Acknowledgments} KJ thanks Maxim Olshanii for helpful discussions. This work was supported by the NSF under Project Nos. PHY-0902906 and PHY-1005571. The large matrix diagonalizations were performed on the large memory nodes of Prof.\ Daniel Steck's parallel cluster at the University of Oregon and the Oregon Center for Optics, funded by the National Science Foundation under Project No.\ PHY-0547926.  


%

\end{document}